\documentclass[runningheads]{svmult}
\usepackage{makeidx}   
\usepackage{graphicx}  
\usepackage{subeqnar}  
\usepackage{multicol}  
\usepackage{physprbb}  
\makeindex             

\begin{document}
\title*{Concepts and methods in the theory of open quantum
        systems}
\toctitle{Concepts and methods in the theory of open quantum
          systems}
%
%
\titlerunning{Open quantum systems}
%
\author{Heinz-Peter Breuer\inst{1}\inst{2}
\and Francesco Petruccione\inst{2}\inst{3}
}
\authorrunning{Heinz-Peter Breuer and Francesco Petruccione}
%
%
\institute{Carl von Ossietzky Universit\"at, Fachbereich Physik,
  D-26111 Oldenburg
\and Physikalisches Institut, Universit\"at Freiburg,
Hermann-Herder-Str. 3, D-79104 Freiburg i. Br.
\and Istituto Italiano per gli Studi Filosofici, Via Monte di Dio 14,
I-80132 Napoli}

\maketitle              

\begin{abstract}
The central physical concepts and mathematical techniques used in
the theory of open quantum systems are reviewed. Particular
emphasis is laid on the interrelations of apparently different
approaches. Starting from the appropriate characterization of the
quantum statistical ensembles naturally arising in the description
of open quantum systems, the corresponding dynamical evolution
equations are derived for the Markovian as well as for the
non-Markovian case.
\end{abstract}

\section{Introduction}
Perfect isolation of quantum systems is not possible since any 
realistic system is influenced by the coupling to an environment,
which typically has a large number of degrees of freedom. A prototypical 
physical system illustrating this situation is given
by an atom interacting with the surrounding radiation field 
\cite{Cohen}. 

In general, a complete microscopic description of the degrees of freedom
of the environment is too complicated. Hence, one has to look for more
simple descriptions of the dynamics of the open system. In principle, 
one should investigate the unitary dynamics of the total system, i.e. 
system and environment, to obtain informations about the reduced system of 
interest by averaging the appropriate observables over the degrees
of freedom of the environment. This is the main concern of the
theory of open quantum systems\index{open quantum system} \cite{TheWork}.

Applications of the theory of open quantum systems are found in almost 
all areas of physics, ranging from quantum optics \cite{WallsMilburn} 
to solid state physics \cite{Weiss},
from chemical physics \cite{Pechukas}  to nanotechnology \cite{Kulik}, from 
quantum information \cite{Nielsen} to 
spintronics \cite{Loss}. On a more fundamental level, the theory of 
open quantum systems 
is relevant for quantum measurement theory \cite{Braginsky} 
and for decoherence and 
the emergence of classicality \cite{Giulini}. 

Usually,  the dynamics of an open quantum system is described in terms 
of the reduced density\index{reduced density matrix} operator which
is obtained from the density operator
of the total system by tracing over the variables of the environment.
In order to eliminate the degrees of freedom of the
environment various approximations are needed which lead to a 
closed equation of motion for the density matrix of the open system. 
The most famous one being the Markov approximation\index{Markov approximation} which 
eventually leads to a so-called quantum master equation\index{quantum master equation}
which, in turn, 
generates a quantum dynamical semi-group\index{quantum dynamical semi-group} 
in the space of  reduced density 
matrices \cite{Alicki}. Prominent representants of such equations are the quantum optical
master equation\index{quantum optical master equation}, and, derived under
slightly different assumptions,
the quantum Brownian motion master equation\index{quantum Brownian motion master equation}, 
with applications in condensed matter physics.

For stronger couplings between the system and the environment
the finite relaxation time of the environment may significantly influence
the dynamics of the open system, making necessary a non-Markovian treatment 
of the reduced system dynamics. Typical examples of such a behaviour
arise in condensed matter physics at low temperatures \cite{Weiss}. 
There are several 
strategies to go beyond a Markovian description of
open quantum systems. Prominent approaches are provided by the influence functional
technique\index{influence functional technique} \cite{Weiss}, 
the cumulant expansion\index{cumulant expansion} \cite{Royer1,Royer2,VanKampen} 
and projection operator techniques\index{projection operator technique} 
\cite{Nakajima,Zwanzig,Chaturvedi}.

During the past  decade another approach to the description of open 
quantum systems has emerged, essentially motivated by the 
experimental evidence for quantum jumps\index{quantum jump} in simple three-level ions \cite{Cook} 
and 
similarly in the photon excitation number of a single mode of a high-Q
cavity \cite{Walther}. The common feature 
of these experiments is that they can be described within the theory of continuous 
measurements\index{continuous measurement} \cite{Braginsky}. Under the condition that the information 
extracted from the 
system by the environment is recoverable, alternative measurement
schemes give rise to different stochastic processes for the
wave function of the open system \cite{Carmichael}. For example,  
the direct photon 
counting of the light emitted by an excited atom gives rise to a piecewise
deterministic process for the wave function of the atom. On this
\textit{selective}\index{selective measurement} level of description 
of open quantum systems the 
typical stochastic processes arising are Markovian. The relation between the 
\textit{selective} wave function approach and the 
\textit{non-selective}\index{non-selective measurement} 
density matrix approach is simply stated \cite{TheWork}: The covariance of the 
stochastic wave function is the density matrix of the reduced system.

\begin{figure}[t]
\begin{center}
\includegraphics[width=0.85\textwidth, angle=180]{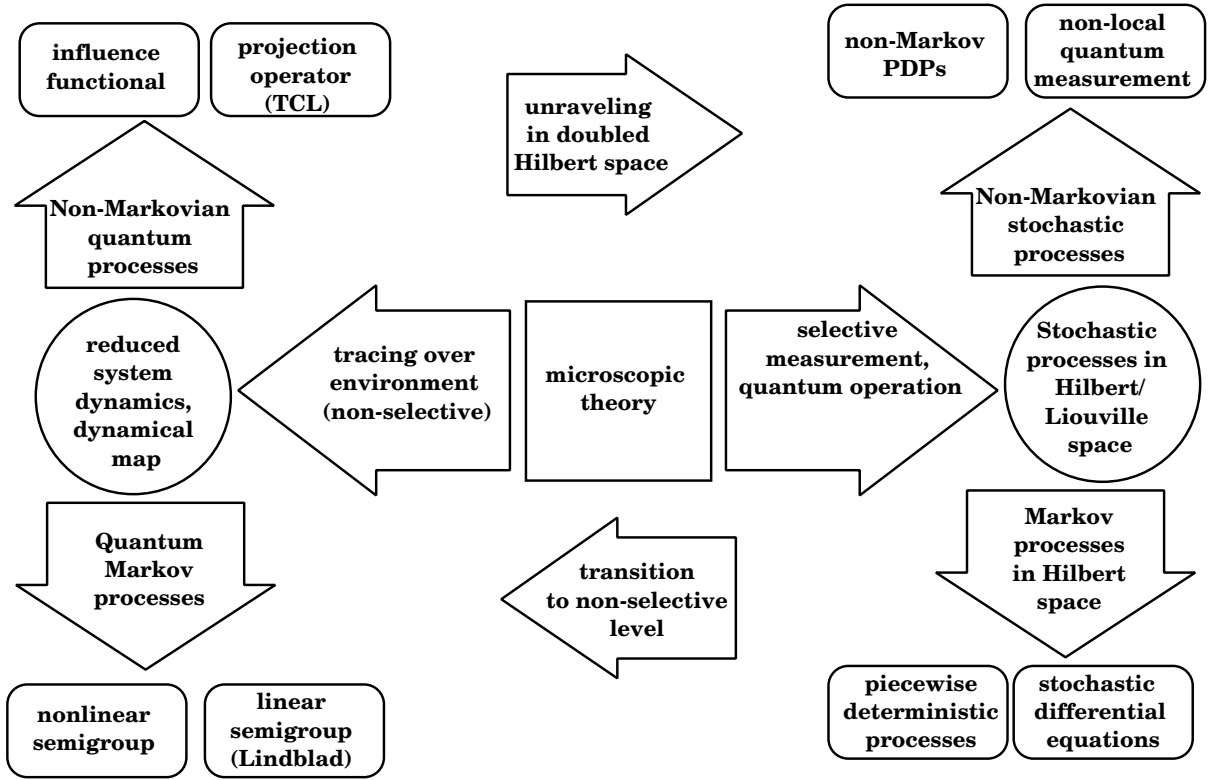}
\end{center}
\caption[]{Basic concepts and methods used in the
 theory of open quantum systems.} \label{fig:MicroscopicTheory}
\end{figure}

The aim of the paper is to guide the reader in a systematic way through 
the different levels of description available for the investigation of 
the dynamics of open quantum systems: 
Density matrix and wave function approaches, 
selective and non-selective measurements,  Markovian and non-Markovian 
approximations, quantum optical and  quantum Brownian motion master equations,
linear and non-linear semigroups, piecewise deterministic and 
diffusive processes for the wave function. Particular emphasis will be laid on establishing 
relationships between apparently different approaches. For this reason
the paper concentrates on the explanation of the flow diagram
Fig. \ref{fig:MicroscopicTheory} which summarizes in a schematic and 
simplified way the main aspects of  the theory of  open quantum systems.
The structure of the paper reflects the main paths through the 
flow diagram Fig. \ref{fig:MicroscopicTheory}. In Sec. 2 we discuss
the centre of the figure, i.e. the microscopic description of the
dynamics of open systems. In Sec. 3 we proceed anti-clockwise from the centre of the
figure to the left, and show how dynamical maps and quantum Markov
processes arise. In Sec. 4 we take the opposite path (clockwise from
the centre to the right) and discuss quantum operations\index{quantum operation}, continuous
measurements\index{continuous measurement} and stochastic processes in
Hilbert space\index{stochastic processes in Hilbert space}. Finally, in
Sec. 5 we describe the upper part of the flow diagram, i.e. the
non-Markovian theory, proceeding clockwise from the centre to the 
left through the upper half of the figure.

\section{Dynamics of open systems: Microscopic theory}
In general terms, an open quantum system is a quantum mechanical
system $S$ with Hilbert space ${\cal{H}}_S$ which is coupled to
another quantum system $B$, the environment, with Hilbert space
${\cal{H}}_B$. Thus, $S$ is a subsystem of the total system $S+B$
living in the tensor product space ${\cal{H}}_S \otimes
{\cal{H}}_B$. Sometimes, the surroundings $B$ of the open system
are termed reservoir to denote an environment with an infinite
number of degrees of freedom. If the reservoir is in thermal
equilibrium one speaks of a heat bath.

Let us denote by $H_S$ the Hamiltonian of the open system, by
$H_B$ the free Hamiltonian of the environment, and by $H_I$ the
Hamiltonian describing the interaction between the system and the
environment. The Hamiltonian of the total system can then be
written as
\begin{equation} \label{TotalHamiltonian}
 H =  H_S \otimes I_B + I_S \otimes H_B + \alpha H_I,
\end{equation}
where $I_S$ and $I_B$ denote the identities in the Hilbert spaces
of the system and of the environment, respectively, and $\alpha$
is a coupling constant. The dynamics of the coupled system
$S+B$ is thus assumed to be Hamiltonian.

An open system $S$ is singled out by the fact that all observables
$A$ of interest refer to this system. Such observables are of the
form $A \otimes I_B$, where $A$ acts in the Hilbert space of the
open system. If the state of the total system $S+B$ is described
by some density matrix $\rho$, then the expectation value of the
observable $A$ is determined by
\begin{equation}
\langle A \rangle = \mathrm{tr}_S \{A \rho_S \},
\end{equation}
where
\begin{equation}
 \rho_S = \mathrm{tr}_B \rho
\end{equation}
is the reduced density matrix\index{reduced density matrix}. In the above equations
$\mathrm{tr}_S$ and $\mathrm{tr}_B$ denote, respectively, the
partial traces over the degrees of freedom of the open system $S$
and of the environment $B$. The reduced density matrix is
of central interest for the theory of open quantum systems. As we
mentioned, the total density matrix evolves unitarily and, hence,
the time-development of the reduced density matrix may be
represented in the form
\begin{equation} \label{EVOL}
 \rho_S(t) = \mathrm{tr}_B
 \left\{ U(t,0) \rho(0) U^{\dagger}(t,0)\right\},
\end{equation}
where the initial state of the total system at time $t=0$ is given
by $\rho(0)$ and $U(t,0)$ is the time-evolution operator of the
total system over the time interval from $t=0$ to $t >0$. The
corresponding differential form of the evolution is obtained from
a partial trace over the environment of the von Neumann
equation\index{von Neumann equation},
\begin{equation}
 \frac{d}{dt} \rho_S(t) = - i \mathrm{tr}_B [H(t), \rho(t)].
\end{equation}
In the following we will survey the most important approaches and
approximations to the above exact equation of motion. 
To this end we will first describe the paths starting from the centre
of the flow diagram Fig. \ref{fig:MicroscopicTheory} to the left and
to the right through the lower part of the figure.

\section{Dynamical maps and quantum Markov processes}

\begin{figure}[t]
\begin{center}
\includegraphics[width=0.85\textwidth]{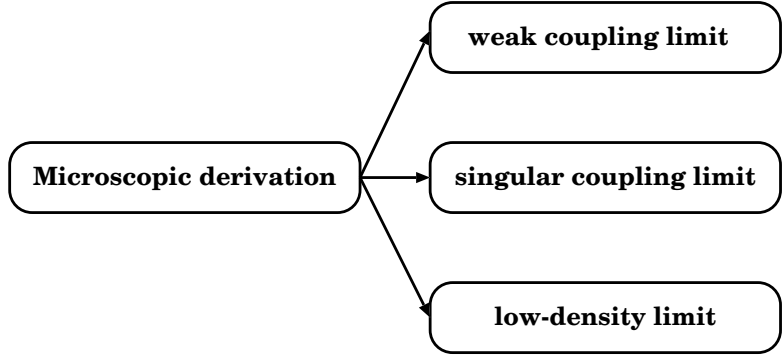}
\end{center}
\caption[]{Schematic overview of the cases in which a Markovian
 quantum master equation can be derived from an underlying
 microscopic theory.} \label{fig:MicroDerivLimits}
\end{figure}

Given that the initial state is of the form
$\rho(0)=\rho_S(0)\otimes\rho_B$, the dynamics expressed through
Eq.~(\ref{EVOL}) can be viewed as a map of the state space of the
reduced system which maps the initial state $\rho_S(0)$ to the
state $\rho_S(t)$ at time $t\geq 0$,
\begin{equation} \label{DYNMAP}
 \rho_S(0) \mapsto \rho_S(t) = V(t) \rho_S(0).
\end{equation}
For a fixed $t$ this map is known as dynamical map\index{dynamical map}. Considered as
a function of time $t$ it provides a one-parameter family
$\{V(t)|t\ge 0\}$ of dynamical maps. If the characteristic time
scale over which the reservoir correlation functions decay are
much smaller than the characteristic time scale of the system's
relaxation, it is justified to neglect memory effects in the
reduced system dynamics and one expects a Markovian type
behaviour, which may be formalized by the semi-group property
\begin{equation}
 V(t_1)V(t_2) = V(t_1+t_2), \quad t_1,t_2 \ge 0.
\end{equation}
The one-parameter family of dynamical maps then becomes a quantum
dynamical semi-group\index{quantum dynamical semi-group}. Introducing the corresponding 
generator $\mathcal{L}$ one immediately obtains an equation of motion for the
reduced density matrix of the open system of the form
\begin{equation} \label{LINDBALD}
 \frac{d}{dt} \rho_S(t) = \mathcal{L} \rho_S(t).
\end{equation}
Such an equation is called a Markovian quantum master equation.
The most general form of the generator $\mathcal{L}$ is provided
by a theorem due to Gorini, Kossakowski and Sudarshan \cite{Gorini} and by a
theorem of Lindblad \cite{Lindblad} according to which
\begin{equation}  \label{LINDBLAD-GENERATOR}
 \mathcal{L} \rho_S = -i[H,\rho_S]
 + \sum_i \gamma_i \left( A_i \rho_S A^{\dagger}_i
 - \frac{1}{2} A^{\dagger}_i A_i \rho_S - \frac{1}{2} \rho_S
 A^{\dagger}_i A_i \right).
\end{equation}
Here, $H$ is the generator of the coherent part of the evolution
(which need not be identical to the free Hamiltonian of the
system) and the $A_i$ are  system operators with corresponding
relaxation times $\gamma_i$. Eq.~(\ref{LINDBALD}) with this form
of the generator is often referred to as Lindblad
equation\index{Lindblad equation}. In a
number of physical situations a quantum master equation whose
generator is exactly of the Lindblad form can be derived from the
underlying microscopic theory under certain approximations. The
most important cases are the weak-coupling limit\index{weak-coupling limit}, the singular
coupling limit\index{singular coupling limit} and the low density
limit\index{low density limit} (see
Fig.~\ref{fig:MicroDerivLimits}).

\begin{figure}[t]
\begin{center}
\includegraphics[width=0.85\textwidth]{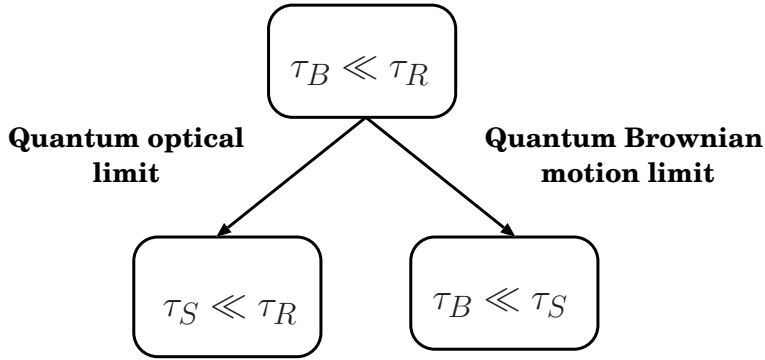}
\end{center}
\caption[]{A comparison of the typical time scales arising in the
  microscopic derivations of the quantum
  optical and the quantum Brownian motion limit of the quantum master
  equation. $\tau_B$ denotes the correlation time of the environment,
  $\tau_R$ the relaxation time of the open system, and $\tau_S$ the
  time scale of its systematic evolution.}
\label{fig:OpticalBromoLimit}
\end{figure}

A widely used Markovian quantum master equation that can be
derived from a microscopic Hamiltonian for the total system in the
weak-coupling limit is the quantum optical master
equation\index{quantum optical master equation} \cite{WallsMilburn}. The
central assumption underlying the weak-coupling approximation is
that the times scales of the system's relaxation $\tau_R$ and of
the correlation time of the environment $\tau_B$ are clearly
separated, i.e. $\tau_B \ll \tau_R$. One further essential
approximation entering the derivation of the quantum optical
master equation is the so-called rotating wave
approximation\index{rotating wave approximation}. The
physical condition behind this approximation is the following one:
The time scale $\tau_S$ of the systematic evolution of the reduced
system is small compared to its relaxation time $\tau_R$, i.e.
$\tau_S \ll \tau_R$. Unfortunately, this condition is violated in
many physical applications involving stronger couplings and low
temperatures. It occurs that the systematic dynamics of the
reduced system may even by slow compared to the correlation time
of the environment, that is $\tau_B \ll \tau_S$. Such cases lead
to the so-called quantum Brownian motion master 
equation\index{quantum Brownian motion master equation}
\cite{Gardiner} (see
Fig.~\ref{fig:OpticalBromoLimit}).

\begin{figure}[t]
\begin{center}
\includegraphics[width=0.85\textwidth]{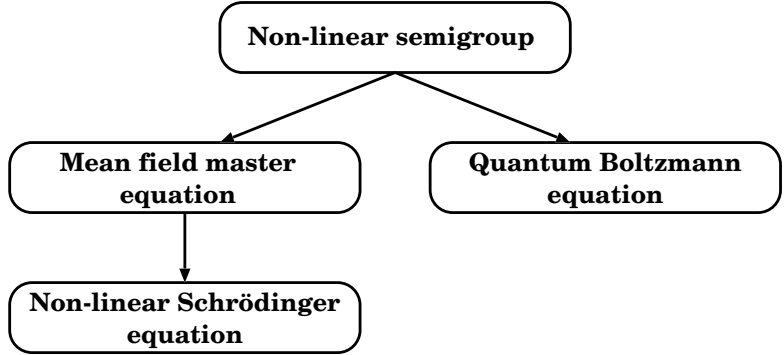}
\end{center}
\caption[]{Schematic overview of typical non-linear semi-groups. A
  typical example for a mean-field equation are the laser equations
  and an example for the application of the non-linear Schr\"odinger equation is super-radiance.}
\label{fig:NonLinearSemiGroup}
\end{figure}

In the treatment of open many-body systems one also encounters
non-linear quantum master equations\index{non-linear quantum master equation} 
for the reduced one-particle
density matrix \cite{AlickiMesser}. In many cases the corresponding generator takes
on the structure of a Lindblad generator whose coefficients depend
parametrically on the density matrix. This means that the
generator $\mathcal{L}=\mathcal{L}[\rho_S]$ provides a
super-operator which represents a function of $\rho_S$ and which
is of Lindblad form for each fixed argument. This immediately
leads to a master equation of the general form
\begin{equation}
 \frac{d}{dt} \rho_S(t) = \mathcal{L}[\rho_S(t)] \rho_S(t).
\end{equation}
Some prominent examples of non-linear quantum master equations are
shown in Fig.~\ref{fig:NonLinearSemiGroup}. At this point we arrived
at the lower left corner of Fig. \ref{fig:MicroscopicTheory}. Now we
go back to the centre of the figure and will explain why it might be
necessary to follow the clockwise path from the centre to the right.

\section{Quantum operations, continuous measurements and stochastic
         processes in Hilbert space}
In order to extract information from a quantum system a
measurement must be carried out. Let us consider the situation
that an open system is measured through an indirect
measurement\index{indirect measurement} on
its environment. Thus, the open system is the quantum object to be
measured, while the environment plays the role of the quantum
probe. The latter is measured by means of a classical apparatus,
once correlations between object and probe system are created as a
result of their interaction.

Let us assume that the interaction between the open system and the
environment begins at time $t=0$. At this time the state of the
system is characterized by the density matrix $\rho_S$ and the
state of the environment, the quantum probe, is given by
\begin{equation}
 \rho_B = \sum_k p_k |\phi_k \rangle \langle \phi_k|.
\end{equation}
The initial density matrix of the total system is thus $\rho(0) =
\rho_S \otimes \rho_B$. At time $\tau$ a classical measurement
apparatus measures the bath observable
\begin{equation}
 R = \sum_m r_m |\varphi_m\rangle \langle \varphi_m|,
\end{equation}
the spectrum of which is assumed to be discrete and
non-degenerate, for simplicity. The application of the von
Neumann-L\"uders projection postulate\index{von Neumann-L\"uders
  projection postulate} to the classical measuring
device shows that the reduced system's state after the measurement
is given by
\begin{equation} \label{IndirectSelectiveMeasure}
 \rho'_m= P(m)^{-1} \Phi_m(\rho_S),
\end{equation}
where
\begin{equation} \label{eq:operation}
 \Phi_m(\rho_S) = \sum_k \Omega_{mk} \rho_S \Omega^{\dagger}_{mk},
\end{equation}
with
\begin{equation}
 \Omega_{mk} = \sqrt{p_k} \langle \varphi_m|U|\phi_k\rangle,
\end{equation}
and
\begin{equation}
 P(m) = \mathrm{tr}_S \{ \Phi_m(\rho_S) \}.
\end{equation}
In these formulae $U$ is the time-development operator of the
total system describing the coupled evolution of quantum system
and quantum probe. The probability that the measurement outcome is
$r_m$ is given by $P(m)$. Under the condition that the outcome is
$r_m$, the corresponding transformation of the reduced system's
density matrix from the initial state $\rho_S$ to the new state
$\rho'_m$ is provided by the map $\Phi_m(\rho_S)$ which is known
as a quantum operation\index{quantum operation}. Like the dynamical map (\ref{DYNMAP}) it
represents a convex-linear and completely positive map. The
representation (\ref{eq:operation}) in terms of the operators
$\Omega_{mk}$ takes on the general form required by the
representation theorem of quantum operations \cite{Krauss}.

When writing Eq.~(\ref{IndirectSelectiveMeasure}) it is assumed
that the measurement is a selective\index{selective measurement} one, i.e. that after the
measurement the information on the measurement outcomes is
retained. Therefore, the original ensemble described by $\rho_S$
is split into a number of sub-ensembles described by the various
$\rho'_m$, each sub-ensemble being conditioned on a specific
outcome $r_m$. By contrast, in the case of a non-selective
measurement\index{non-selective measurement} the final state of the system after the measurement is
given by
\begin{equation}
 \rho' = \sum_m P(m) \rho'_m = \sum_m \Phi_m(\rho_S),
\end{equation}
which describes the ensemble obtained after re-mixing the
sub-ensembles after the measurement. Thus, the first important
step in the description of an open quantum system is to make
clarity about the type of measurement which is performed in order
to obtain information about the system. In other words, we have to
know whether the (indirect) measurement is selective or not. The
microscopic theory of the total (closed) system has to be
supplemented with this information as is indicated in
Fig.~\ref{fig:MicroscopicTheory}. This important distinction has
fundamental consequences on the choice of the formalism required
to describe the open system. In fact, selective and non-selective
measurements require different characterizations of the quantum
statistical properties of the open system.

Let us now follow another important path through the flow diagram.
Again we start at the centre of Fig.~\ref{fig:MicroscopicTheory}
and proceed clockwise to the right into the lower part of the
diagram. The decisive aspect of this path is the indirect
selective measurement of the open quantum system, which requires a
particular characterization of the quantum statistical ensembles.

A quantum statistical ensemble characterized in terms of a density
matrix $\rho$ describes a disordered set of a large number of
individual quantum systems, where each system has been prepared in
one of a certain set of states $\psi_{\alpha}$. The preparation
measurements could have been carried out, for example, through the
measurements of complete sets of commuting observables. The
various $\psi_{\alpha}$, however, need not be orthogonal. A
mixture of these states with respective weights $w_{\alpha}$ gives
rise to an ensemble $\mathcal{E}_{\rho}$ which is described by the
density matrix
\begin{equation}
 \rho = \sum_{\alpha} w_{\alpha}
 |\psi_{\alpha}\rangle\langle\psi_{\alpha}|.
\end{equation}

\begin{figure}[t]
\begin{center}
\includegraphics[width=0.8\textwidth]{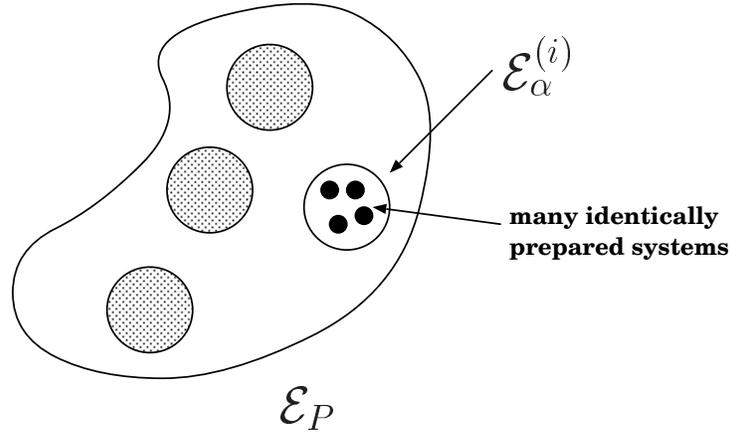}
\end{center}
\caption[]{Schematic representation of an ensemble of type $\mathcal{E}_P$.}
\label{fig:EpsilonP}
\end{figure}

A different kind of ensembles which is appropriate for the
description of selective measurements and which will be denoted by
${\mathcal{E}}_P$ is obtained as follows. Consider a collection of
pure quantum statistical ensembles ${\mathcal{E}}_{\alpha}$
describable by corresponding states $\psi_{\alpha}$. We want to keep, as
required for a selective measurement, the information that a
particular quantum system belongs to a particular ensemble
${\mathcal{E}}_{\alpha}$. For this purpose we take $N_{\alpha}$
identically prepared copies
${\mathcal{E}}^{(1)}_{\alpha},{\mathcal{E}}^{(2)}_{\alpha},\ldots,
{\mathcal{E}}^{(N_{\alpha})}_{\alpha}$ of ${\mathcal{E}}_{\alpha}$
for each ${\alpha}$. The new ensemble ${\mathcal{E}}_P$ is then
the collection of these ensembles, that is an {\em{ensemble of
ensembles}}\index{ensemble of ensembles}:
\begin{equation} \label{eq:EnsembleEnsamble}
{\mathcal{E}}_P = \left\{ {\mathcal{E}}^{(1)}_1,\ldots,
{\mathcal{E}}^{(N_1)}_1, {\mathcal{E}}^{(1)}_2,\ldots,
{\mathcal{E}}^{(N_2)}_2,\ldots,
{\mathcal{E}}^{(1)}_{\alpha},\ldots,
{\mathcal{E}}^{(N_{\alpha})}_{\alpha},\ldots \right\}.
\end{equation}
The numbers $N_{\alpha}$ are chosen such that
$w_{\alpha}=N_{\alpha}/N$, where $N=\sum_{\alpha}N_{\alpha}$,
which implies that ${\mathcal{E}}_{\alpha}$ appears with the
statistical weight $w_{\alpha}$ in the set
(\ref{eq:EnsembleEnsamble}). The decisive distinction of an
${\mathcal{E}}_P$-ensemble to an $\mathcal{E}_{\rho}$-ensemble is
that ${\mathcal{E}}_P$ represents a set whose elements are again
sets, namely the ensembles ${\mathcal{E}}^{(i)}_{\alpha}$. A
schematic picture of an ensemble of type $\mathcal{E}_P$ is given
in Fig.~\ref{fig:EpsilonP}.

Since the various ${\mathcal{E}}_{\alpha}$ making up
${\mathcal{E}}_P$ are represented by their corresponding state
vector $\psi_{\alpha}$, an ensemble of the type ${\mathcal{E}}_P$
gives rise to a probability distribution\index{probability distribution on Hilbert space} 
$P=P[\psi]$ on Hilbert
space ${\mathcal{H}}_S$. Consequently, the state vector becomes a
random vector in Hilbert space (see \cite{TheWork} and references therein). 
More precisely speaking, the
functional $P[\psi]$ represents a probability distribution on
projective Hilbert space, that is on the space of rays in
${\mathcal{H}}_S$. Regarded as a density functional on
$\mathcal{H}_S$, the distribution $P[\psi]$ is therefore subjected
to the normalization condition
\begin{equation}
 \int D\psi D\psi^{\ast} P[\psi] =1,
\end{equation}
with an appropriate volume element $D\psi D\psi^{\ast}$ in Hilbert
space, it vanishes outside the unit sphere $||\psi||=1$ in Hilbert
space, and it is invariant under changes of the phase of the state
vector, that is $P[\exp(i \chi) \psi ] = P[\psi]$. The density
matrix $\rho$ characterizing the corresponding ensemble
$\mathcal{E}_{\rho}$ now appears as the covariance matrix of the
random state vector $\psi$ which is defined as the expectation
value E of the quantity $|\psi\rangle\langle\psi|$. In terms of
the probability density functional $P[\psi]$ we have
\begin{equation} \label{eq:RhoDefinition}
 \rho = {\rm{E}} \left[ |\psi\rangle\langle\psi| \right]
 \equiv \int D\psi D\psi^{\ast} P[\psi] | \psi \rangle \langle \psi|.
\end{equation}
This relation clearly reveals that the statistics of an ensemble
$\mathcal{E}_{\rho}$ is completely determined by the probability
density $P[\psi]$, while the converse is obviously not true.

\begin{figure}[t]
\begin{center}
\includegraphics[width=\textwidth]{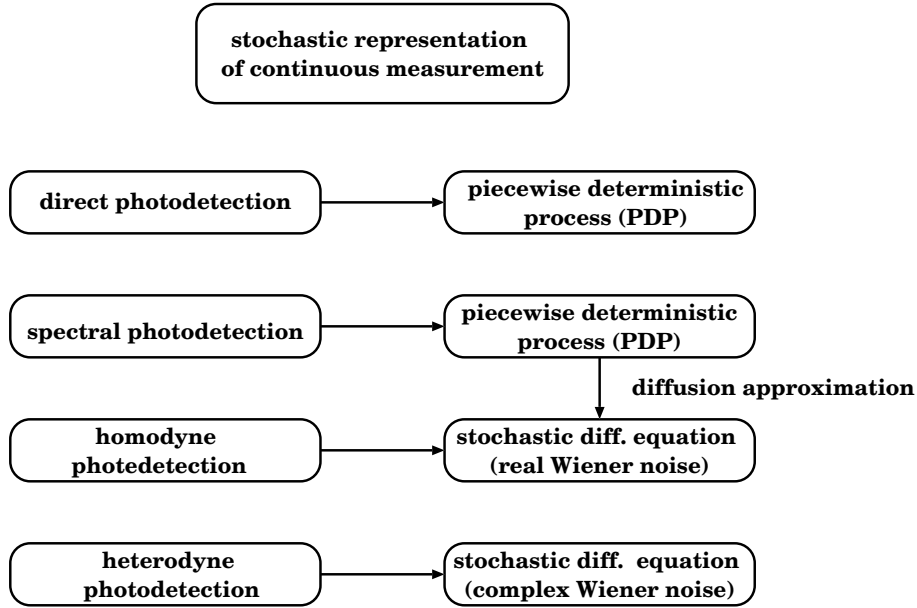}
\end{center}
\caption[]{Different continuous measurement schemes give rise to
  different Markovian stochastic processes for the wave function of
  the reduced system. The stochastic Schr\"odinger equations are
  obtained as  diffusion limits of appropriate piecewise deterministic processes.}
\label{fig:StochReprContMeasure}
\end{figure}

The formalism of probability densities on Hilbert space now allows
the description of the dynamics of an open system which is
continuously monitored through its environment (see the lower part
of the flow diagram shown in Fig.~\ref{fig:MicroscopicTheory}). The probability
density then becomes a time-dependent functional $P=P[\psi,t]$ and
the reduced system's state vector provides a stochastic process
$\psi(t)$ in Hilbert space. Physically, $\psi(t)$ represents the
state of the reduced system which is conditioned on a specific
readout of the measurement carried out on the environment.
Consequently, the stochastic evolution depends on the measurement
scheme used to monitor the environment. As an example we write a
stochastic differential equation corresponding to the Lindblad
equation (\ref{LINDBALD}) with generator
(\ref{LINDBLAD-GENERATOR}):
\begin{equation} \label{PDP-SDGL}
 d\psi(t) = -i G(\psi(t)) dt + \sum_i
 \left( \frac{A_i \psi(t)}{||A_i\psi(t)||} - \psi(t) \right)
 dN_i(t).
\end{equation}
A stochastic differential equation of this form is known as a
piecewise deterministic process\index{piecewise deterministic process}. 
The first term on the right-hand
side describes the deterministic evolution periods given by the
non-linear Schr\"odinger equation\index{non-linear Schr\"odinger equation}
\begin{equation} \label{NONLINEAR_SCHROEDINGER}
 \frac{d}{d t} \psi = -i G(\psi) \equiv H\psi  - \frac{i}{2} \sum_{i}
 \gamma_i \left( A_i^{\dagger} A_i - || A_i \psi ||^2 \right)\psi.
\end{equation}
The deterministic pieces of the motion are interrupted by
instantaneous changes of the state vector according to $\psi
\longrightarrow A_i\psi/||A_i\psi||$. These so-called quantum
jumps\index{quantum jump} are described by the second term on the right-hand side of
Eq~(\ref{PDP-SDGL}) which involves the random numbers $dN_i(t)$.
These numbers take on the values $0$ or $1$ and represent
independent Poisson increments with the expectation values
\begin{equation} \label{POISSON-INCR}
 {\rm{E}}\left[ dN_i(t) \right] = \gamma_i ||A_i\psi(t)||^2 dt.
\end{equation}
A stochastic evolution equation of this form is obtained, for
example, if the reduced system represents an excited atom, the
environment is an electromagnetic field vacuum and the system's
state is monitored through the direct observation of the emitted
quanta.

Depending on the measurement scheme, it may be appropriate to
perform a diffusion approximation\index{diffusion approximation} of the above piecewise
deterministic process. This leads to a diffusion process for the
wave function of the reduced system involving real or complex
Wiener noise. An overview of the Markov processes for the wave
function of the open system related to a certain continuous
measurement\index{continuous measurement} scheme is given in
Fig.~\ref{fig:StochReprContMeasure}.

\section{Non-Markovian processes}
Let us now turn our attention to the description of the
non-Markovian\index{non-Markovian process} dynamics of an open quantum system. We proceed
counter clockwise as indicated in the upper half of the flow
diagram shown in Fig.~\ref{fig:MicroscopicTheory}. Again, the
starting point is provided by an underlying microscopic theory for
the total system. Unlike in the Markovian case, the first step
consists in deriving an exact equation of motion for the reduced
density matrix of the open system through an elimination of the
dynamical variables of the environment. Of course, a closed
analytical expression for the reduced density matrix can be
obtained only for a few analytically solvable models, such as for
the damped harmonic oscillator and for free Brownian motion
\cite{FeynmanVernon,CALDEIRA,GRABERT}. In many interesting cases,
however, exact representations for the reduced density matrix
serve as starting points of a perturbation expansion in the
system-environment coupling and of the development of numerical
integration schemes.

\begin{figure}[t]
\begin{center}
\includegraphics[width=\textwidth]{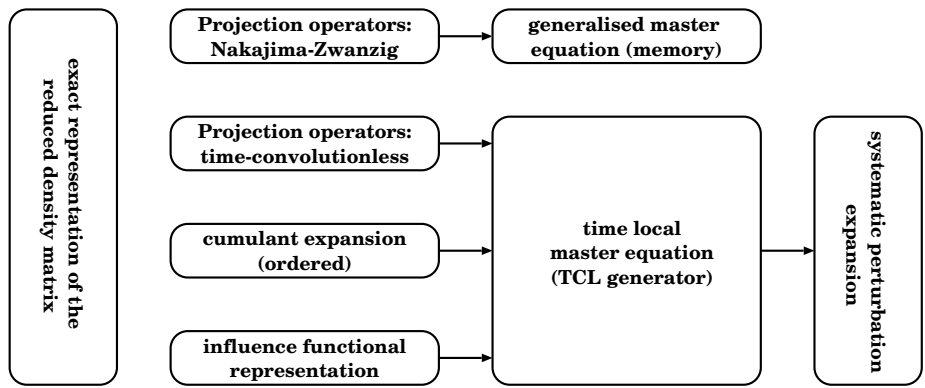}
\end{center}
\caption[]{Schematic overview of the different approaches to the
  description of the non-Markovian dynamics of open quantum systems.}
\label{fig:NonMarkovianApproaches}
\end{figure}

As indicated in Fig. \ref{fig:MicroscopicTheory} and in more
detail in Fig. \ref{fig:NonMarkovianApproaches} there are several
possibilities to carry out this program. Probably the most
prominent one is to use the Nakajima-Zwanzig projection operator
technique \cite{Nakajima,Zwanzig}. In this approach one derives a formally exact equation
for the reduced density matrix of the open system in the form of
an integro-differential equation involving a certain memory
kernel. The practical disadvantage of this approach is that the
perturbation expansion of the memory kernel does not simplify the
convolution structure of the equations whose numerical solution
could be quite involved (see also \cite{Royer2}).

As far as a numerical and/or perturbation approach is concerned a
more appropriate strategy is to derive a time-local master
equation\index{time-local master equation} for the open system's density matrix $\rho_S(t)$ which
takes the form
\begin{equation} \label{TCL-MASTER}
 \frac{d}{dt} \rho_S(t) = {\mathcal{K}}(t) \rho_S(t).
\end{equation}
${\mathcal{K}}(t)$ is a time-dependent generator, a super-operator
in the reduced system's Hilbert space ${\mathcal{H}}_S$. Employing
the time-convolutionless (TCL) projection operator technique
\cite{Shibata1,Shibata2} one can show that a master equation of
the form (\ref{TCL-MASTER}) indeed exists for small and
intermediate couplings in the case of factorizing initial
conditions. We remark that a similar method can also be used in
the general case of a correlated initial state, in which case the
time-local master equation contains an additional inhomogeneous
term. It should be noted that Eq.~(\ref{TCL-MASTER}) is local in
time, i.~e. that it does not involve an integration over the past
history of the reduced system. Due to the explicit time-dependence
of the TCL generator ${\mathcal{K}}(t)$, however, it does not lead
to a quantum dynamical semigroup and, therefore, the generator
need not be in Lindblad form.

Eq.~(\ref{TCL-MASTER}) provides an appropriate starting point for
a perturbation expansion with respect to some coupling parameter
$\alpha$. To this end, one expands the time-local generator in
powers of $\alpha$,
\begin{equation} \label{TCL-EXPANSION}
 {\mathcal{K}}(t) = \sum_{n=1}^{\infty} \alpha^n
 {\mathcal{K}}_n(t),
\end{equation}
and solves the corresponding equation of motion which is obtained in the
desired order in $\alpha$ either analytically or numerically. Such
an expansion may be found in two different ways. One way is to
start from the formal solution of the von Neumann equation of the
total system and to use an expansion in terms of ordered
cumulants\index{ordered cumulant}
\cite{Royer1,Royer2,VanKampen}. Another way is to invoke the Feynman-Vernon
influence functional\index{influence functional technique} 
representation of the reduced density matrix
\cite{FeynmanVernon} and to obtain an expansion of the generator
directly in terms of the influence phase. Both strategies, where
applicable, yield identical expansions of the TCL generator
\cite{BMP_MQC2}. The different approaches to the description of
the non-Markovian dynamics of an open system are summarized in
Fig. \ref{fig:NonMarkovianApproaches}.

In general, one expects that a time-local master equation whose
generator consists of only the first few terms of the expansion
provides a good description of the reduced dynamics for weak and
moderate couplings. However, it should be emphasized that an
expansion of the form (\ref{TCL-EXPANSION}) need not exist for
strong couplings\index{strong coupling}. What happens in these cases is that the initial
state $\rho_S(0)$ is not uniquely determined by the state
$\rho_S(t)$ at time $t$. Specific examples of the application of
this technique to physical models and of the breakdown of the TCL
expansion in the strong coupling regime are discussed in
\cite{TheWork}.

Within any given order the time-local master equation for the
reduced density matrix $\rho_S(t)$ which results from the above
procedure takes the following form,
\begin{equation} \label{eq:rho_mo}
 \frac{d}{d t}\rho_S(t) = A(t)\rho_S(t)
 + \rho_S(t)B^\dagger(t) + \sum_i C_i(t)\rho_S(t)D_i^\dagger(t),
\end{equation}
with some time-dependent linear operators $A(t)$, $B(t)$, $C_i(t)$
and $D_i(t)$. This equation is linear in $\rho_S(t)$ and local in
time, but it needs not be in Lindblad form. However, it is
important to realize that a stochastic representation of the
dynamics given by this equation is possible \cite{BKP99,TheWork}.

The key point for a stochastic unraveling of the non-Markovian
master equation (\ref{eq:rho_mo}) is the usage of a stochastic
process in the doubled Hilbert 
space\index{stochastic process in doubled Hilbert space} ${\mathcal{H}}_S \oplus
{\mathcal{H}}_S$. This means that we represent the dynamics
through a pair of stochastic wave functions
\begin{equation}
 \theta(t) = \left( \begin{array}{c}
                  \phi(t) \\
                  \psi(t)
                  \end{array} \right),
\end{equation}
such that $\theta(t)$ becomes a stochastic process in the doubled
Hilbert space. An appropriate stochastic differential equation for a
piecewise deterministic process in the doubled Hilbert space is
the following,
\begin{equation}
  \label{eq:gen_un}
  d\theta(t) = -iG(\theta,t)dt
  +\sum_i\left(\frac{\|\theta(t)\|}{\left\|
  J_i(t)\theta(t)\right\|}
  J_i(t)\theta(t)-\theta(t)\right)dN_i(t).
\end{equation}
The numbers $dN_i(t)=0,1$ are again independent Poisson increments
with expectation values
\begin{equation} \label{eq:dbl_mean}
 {\rm{E}}[dN_i(t)] = \frac{\left\| J_i(t)\theta(t)\right\|^2}
 {\|\theta(t)\|^2}dt,
\end{equation}
and the non-linear operator $G(\theta,t)$ is defined as
\begin{equation}
  \label{eq:G_dbl}
   G(\theta,t)=i\left( F(t)+\frac{1}{2}\sum_i
  \frac{\left\| J_i(t)\theta(t)\right\|^2}{\|\theta(t)\|^2}\right)
  \theta(t),
\end{equation}
with time-dependent operators of block structure,
\begin{equation}
  \label{eq:AB}
   F(t)=
  \left(\begin{array}{cc}
  A(t)&0\\
  0&B(t)\end{array}\right),\quad
   J_i(t)=
  \left(\begin{array}{cc}
  C_i(t)&0\\
  0&D_i(t)\end{array}\right).
\end{equation}

With the help of the calculus of piecewise stochastic processes
\cite{TheWork} it
can now be demonstrated that the expectation value
\begin{equation}
 \rho_S(t) = {\rm{E}} \left[ |\phi\rangle\langle\psi| \right]
\end{equation}
satisfies the non-Markovian master equation (\ref{eq:rho_mo}).
Thus, any time-local master equation allows a stochastic
unraveling in the doubled Hilbert space. This fact makes it
possible to use the apparatus of stochastic processes in the
analysis of non-Markovian quantum processes and to design the
corresponding stochastic simulation algorithms. It also enables us
to close, finally, the flow diagram shown in
Fig.~\ref{fig:MicroscopicTheory} through the path following the
upper half of the diagram.

%
 \clearpage
 \addcontentsline{toc}{section}{Index}
 \flushbottom
 \printindex


\begin{thebibliography}{8.}
\addcontentsline{toc}{section}{References}
\bibitem{Cohen} C. Cohen-Tannoudj, J. Dupont-Roc, G. Grynberg,
                \textit{Atom-Photon Interactions} (John Wiley, New York, 1998).
\bibitem{TheWork} H. P. Breuer and F. Petruccione, \textit{The Theory
  of Open Quantum Systems} (Oxford University Press, Oxford, 2002).
\bibitem{WallsMilburn} D. F. Walls and G. J. Milburn, \textit{Quantum Optics}
               (Springer-Verlag, Berlin, 1994).
\bibitem{Weiss} U. Weiss, \textit{Quantum Dissipative Systems}, Volume 2 of
                 \textit{Series in Modern Condensed Matter Physics} 
            (World Scientific, Singapore, 1999).
\bibitem{Pechukas} P. Pechukas and U. Weiss (guest editors), 
               \textit{Quantum Dynamics of Open Systems},
              Chemical Physics, Special Issue, Vol. 268, Nos. 1-3 (2001).
\bibitem{Kulik} I. O. Kulik and R. Ellialtioglu (Eds.),
            \textit{Quantum Mesoscopic Phenomena and Mesoscopic
              Devices in Microelectronics}, NATO Series, Series C:
            Mathematical and Physical Sciences, Vol. 559 (Kluwer
            Academic Publishers, Dordrecht, 2000).
\bibitem{Nielsen} M. A. Nielsen and I. L. Chuang, 
           \textit{Quantum Computation and Quantum Information} 
             (Cambridge University Press, Cambridge, 1997).
\bibitem{Loss}  D.D. Awschalom, D. Loss, and N. Samarth (eds.),  
           \textit{Semiconductor Spintronics and Quantum Computation},
            Series on Nanoscience and Technology, (Springer-Verlag, Berlin,
             2002).
\bibitem{Braginsky} V. B. Braginsky and F. Ya. Khalili, 
           \textit{Quantum Measurement} (Cambridge University Press, 
                     Cambridge, 1992).
\bibitem{Giulini} D. Giulini, E. Joos, C. Kiefer, J. Kupsch, 
                 I.-O.Stamatescu, H. D. Zeh, \textit{Decoherence and the 
             Appearance of a Classical World in Quantum Theory}
              (Springer-Verlag, Berlin, 1996).
\bibitem{Alicki} R. Alicki and M. Fannes, \textit{Quantum Dynamical Systems}
      (Oxford University Press, Oxford, 2001).
\bibitem{Royer1} A. Royer, \textit{Phys. Rev. A}, \textbf{6} (1972) 1741.
\bibitem{Royer2} A. Royer, this volume.
\bibitem{VanKampen} N. G. van Kampen,
 \textit{Physica}, \textbf{74} (1974) 215--238; \textbf{74} (1974) 239--247.
\bibitem{Nakajima} S. Nakajima, Progr. Theor. Phys. \textbf{20} (1958) 948-959.
\bibitem{Zwanzig} R. Zwanzig, J. Chem. Phys. \textbf{33} (1960) 1338-1341.
\bibitem{Chaturvedi} S. Chaturvedi and F. Shibata, Z. Phys. \textbf{B35}
          (1979)     297-308.
\bibitem{Cook} R. J. Cook, Progr. Opt. \textbf{28} (1990) 361.
\bibitem{Walther} G. Benson, G. Raithel, H. Walther, Phys. Rev. Lett.
             \textbf{72} (1994) 3506.
\bibitem{Carmichael} H. Carmichael, \textit{An Open Systems Approach 
            to Quantum Optics}, Lecture Notes in Physics m18 
          (Springer-Verlag, Berlin, 1993).
\bibitem{Gorini} V. Gorini, A. Kossakowski, E. C. G. Sudarshan,
                 J. Math. Phys. \textbf{17} (1976) 821-825.
\bibitem{Lindblad} G. Lindblad, Comm. Math. Phys. \textbf{48} (1976)
  119-130. 
\bibitem{Gardiner} C. W. Gardiner and P. Zoller, \textit{Quantum
    Noise}, 2nd edition (Springer-Verlag, Berlin, 2000).
\bibitem{AlickiMesser} R. Alicki and J. Messer,
  J. Stat. Phys. \textbf{32} (1983) 299-312.
\bibitem{Krauss} K. Kraus, \textit{States, Effects, and Operations}
  (Springer-Verlag, Berlin, 1983).
\bibitem{FeynmanVernon} R. P. Feynman and F. L. Vernon,
 \textit{Ann. Phys. (N. Y.)}, \textbf{24} (1963) 118--173.
\bibitem{CALDEIRA} A. O. Caldeira and A. J. Leggett,
 \textit{Physica}, \textbf{121A} (1983) 587--616.
\bibitem{GRABERT} H. Grabert, P. Schramm and G.-L. Ingold,
 \textit{Phys. Rep.}, \textbf{168} (1988) 115--207.
\bibitem{Shibata1} F. Shibata, Y. Takahashi and N. Hashitume,
           \textit{J. Stat. Phys.}, \textbf{17} (1977) 171--187.
\bibitem{Shibata2} S. Chaturvedi and F. Shibata, \textit{Z. Phys. B},
            \textbf{35} (1979) 297--308.
\bibitem{BMP_MQC2} H. P. Breuer, A. Ma, F. Petruccione, Time-local
  master equations: influence functional and cumulant expansion, in:
       \textit{Quantum Computing and Quantum Bits in Mesoscopic Systems},
    A.J. Leggett, B. Ruggiero, and P. Silvestrini (eds),
        Kluwer Academic/Plenum Publishers, in press.
\bibitem{BKP99} H.P. Breuer, B. Kappler and F. Petruccione,
          \textit{Phys. Rev. A}, \textbf{59} (1999) 1633--1643.
\end{thebibliography}
\end{document}